# Are Effective Leaders Creative?

Liane Gabora

Department of Psychology, University of British Columbia
Okanagan Campus, Arts Building, 333 University Way, Kelowna BC, V1V 1V7, CANADA

**Abstract**

This paper explains in layperson's terms how an agent-based model was used to investigate the widely held belief that creativity is an important component of effective leadership. Creative leadership was found to increase the mean fitness of cultural outputs across an artificial society, but the more creative the followers were, the greater the extent to which the beneficial effect of creative leadership was washed out. Early in a run when the fitness of ideas was low, a form of leadership that entails the highest possible degree of creativity was best for the mean fitness of outputs across the society. As the mean fitness of outputs increased a transition inevitably occurs after which point a less creative style of leadership proved most effective. Implications of these findings are discussed.

It is widely assumed that creativity is an important component of effective leadership. (Bellows, 2004, Basadur, 2004; Puccio, Murdock, & Mance, 2006; Simon, 1986; Sternberg, Kaufman, & Pretz, 2003). This paper explains in layperson's terms how a computer model was used to examine the relationship between creativity and leadership. Although the results we obtained made sense when we thought about them, there were some interesting surprises.

First we studied how leadership affects the effectiveness and diversity of ideas in a society. Second we studied to what extent creativity is desirable in a leader. I'll begin by telling you a bit about the computer model itself, and then explain the experiments.

THE COMPUTER MODEL

The current model's predecessor was called Meme and Variations or MAV (Gabora, 1995). Its name is a pun on the musical form, 'theme and variations'. MAV was the earliest computer program to model culture as an evolutionary process in its own right. MAV was inspired by the genetic algorithm (GA), a search technique that finds solutions to complex problems by generating a 'population' of candidate solutions (through processes akin to mutation and recombination), selecting the best, and repeating until a satisfactory solution is found.

The computer model is composed of an artificial society of agents in a two-dimensional grid-cell world. Agents consist of (1) a neural network, which encodes ideas for actions and detects trends in what constitutes an effective action, and (2) a body, which implements their ideas as actions. The agents can do two things: (1) invent ideas for new actions, and (2) imitate their neighbors' actions. The computer model enables us to investigate what happens to the diversity and effectiveness of actions in the artificial society over successive rounds (called 'iterations') of invention and imitation. Since the ideas in the model are ideas for actions, diversity is measured by counting how many different actions are being implemented by the agents. Evolution in the biological sense is not taking place; the agents neither die nor have offspring. But evolution in the cultural sense is taking place through the generating and sharing of ideas for actions amongst agents, which over time leads to more effective actions.

In MAV, all agents were equally capable of both inventing and imitating. In the latest version of the computer model called EVOC (for EVOlution of Culture), it is possible to vary how likely an agent is to invent versus imitate.

A TYPICAL RUN

Each iteration, every agent has the opportunity to (1) acquire an idea for a new action, either by imitation, copying a neighbor, or by invention, creating one anew, (2) update their knowledge about what constitutes an effective action, and (3) implement a new action. Effectiveness of actions starts out low because initially all agents are just standing still doing nothing. Soon some agent invents an action that has a higher effectiveness than doing nothing, and this action gets imitated, so effectiveness increases. Effectiveness increases further as other ideas get invented, assessed, implemented as actions, and spread through imitation. The diversity of actions initially increases due to the proliferation of new ideas, and then decreases as agents hone in on the fittest actions. Thus MAV successfully models how 'descent with modification' can occur in a cultural context.

LEADERSHIP EXPERIMENTS

These experiments made use of EVOC's 'broadcasting function'. This enables the actions of a particular agent, the leader or 'broadcaster' to be imitated by not just its immediate neighbors (as is normally the case) but any other agent. Thus broadcasting enables the action implemented by a leader to be visible to all the other agents in the artificial society, referred to as followers. In these experiments, societies consisted of one leader and ninety-nine followers. The leader was chosen randomly and broadcasted throughout the 100-iteration run.

In a first set of simulations, the leader was no more or less creative than the followers. We found that the presence of a leader accelerates convergence on optimal ideas, but does so at the cost of consistently reducing the diversity of ideas (Gabora, 2008a,b). In other words, although they find optimal solutions faster, they end up finding fewer of them. This echoes previous simulation findings that when agents can communicate or exchange ideas, leadership can have adverse effects (Gigliotta, Miglino, & Parisi, 2007). The result suggests that in a fast-changing world where diversity of ideas is beneficial because what is effective today may not be effective tomorrow, it may be particularly important to watch out for situations in which leaders pull individuals off their own creative paths.

The goal of the next set of experiments was to investigate how creative versus uncreative leadership affects the effectiveness and diversity of ideas, and how creative leadership is affected by how creative the followers are (Leijnen & Gabora, 2010). There are two ways a leader can be creative in EVOC. The first way has to do with how OFTEN the leader invents; that is, the ratio of iterations it spent inventing versus imitating. When an agent's frequency of invention is at the maximum of 1.0, it invents a new action every iteration. When the frequency of invention is at the minimum of 0.0, the agent never invents new actions; it only imitates its neighbors' actions. We tried many possibilities ranging between these two extremes, for both leader and followers.

What we found was that when the followers are uncreative, the degree of creativity of the leader matters a lot; the effectiveness of ideas across the society as a whole is positively correlated with the frequency of invention of the leader. However, the more creative the followers are, the greater the extent to which the beneficial effect of creative leadership is washed out. When the followers themselves are creative, the degree of creativity of the leader has almost no impact; in this case, the ideas generated by the society increase over the duration of a run at more or less the same pace, no matter how creative the leader is. The results suggest that creativity may be an important quality for a manager of a relatively uncreative team, but not such an important quality for a manager of a creative team.

We then wanted to know whether the decreased diversity associated with the presence of a leader is still observed when leaders are highly creative or highly uncreative compared to followers. We found that while in the early stages of a run, creative leadership is associated with higher cultural diversity, eventually all agents converge on what the leader is doing no matter how creative the leader (Leijnen & Gabora, 2010). That is, in the long run, leadership diminishes cultural diversity regardless of how creative the leader is.

Yet another set of experiments investigated the effect of not how often the leader invents, but how creative the leader's inventions are; that is, the extent to which a newly invented idea differs from its predecessor idea. It turned out that the optimal degree of creative leadership with respect to this second measure of creativity depends on how far along the society is. Early on in a run, when the fitness of ideas is still low, a form of leadership that entails the highest possible degree of creativity (highest rate of change per new idea) is ideal. However, this situation changes as the run progresses, and eventually a transition occurs, after which point a much lower rate of change per idea (approximately 40%) is best. Although once again one must be cautious about extrapolating from the results of such simulations to the real world, this result suggests that a new start-up company benefits most from highly creative leadership, while a more established company, or one that has stabilized on an established product line, may benefit most from a more conservative form of leadership.

One must be cautious about extrapolating from a simple simulation such as this to the real world. For example, real-world creativity is correlated with emotional instability, affective disorders, and substance abuse (Andreason, 1987; Flaherty, 2005; Jamieson, 1993) which presumably would interfere with effective leadership, and which were not incorporated in these simulations. Moreover, the agents' neural networks are so small that creative novelty is generated does not involve noticing and refining new kinds of connections the way it happens in real minds (Gabora, 2000). Finally, it is also worth noting though that in this artificial world, unlike the real world, agents had only one task to accomplish. Further experiments will investigate whether this result hold true when there are multiple tasks to be accomplished. However, the results of these computer simulations are provocative, and inspire new ways of thinking about the relationship between creativity and leadership. They suggest that the relationship between leadership and creativity is more complex than previously thought.

Cogsci2010LeijnenGabora.pdf